\documentclass[manuscript]{acmart}

\usepackage{multicol}
\usepackage{booktabs}
\usepackage{multirow}
\usepackage{siunitx}

\AtBeginDocument{ \providecommand\BibTeX{{ \normalfont B\kern-0.5em{\scshape i\kern-0.25em b}\kern-0.8em\TeX}}}

\newcommand\blfootnote[1]{%
  \begingroup
  \renewcommand\thefootnote{}\footnote{#1}%
  \addtocounter{footnote}{-1}%
  \endgroup
}

\copyrightyear{2023}
\acmYear{2023}
\setcopyright{rightsretained}
\acmConference[RecSys '23]{Seventeenth ACM Conference on Recommender Systems}{September 18--22, 2023}{Singapore, Singapore}
\acmBooktitle{Seventeenth ACM Conference on Recommender Systems (RecSys '23), September 18--22, 2023, Singapore, Singapore}\acmDOI{10.1145/3604915.3610250}
\acmISBN{979-8-4007-0241-9/23/09}

\begin{document}

\title[Enhancing Large-Scale Recommender Systems with Context-Based Prediction Models]{Unleash the Power of Context: Enhancing Large-Scale Recommender Systems with Context-Based Prediction Models}

\author{Jan Hartman}
\email{jhartman@outbrain.com}
\affiliation{
  \institution{Outbrain}
  \country{Slovenia}
  \city{Ljubljana}
}

\author{Assaf Klein}
\email{aklein@outbrain.com}
\affiliation{
  \institution{Outbrain}
  \country{Israel}
  \city{Netanya}
}

\author{Davorin Kopi\v{c}}
\email{dkopic@outbrain.com}
\affiliation{
  \institution{Outbrain}
  \country{Slovenia}
  \city{Ljubljana}
}

\author{Natalia Silberstein}
\email{nsilberstein@outbrain.com}
\affiliation{
  \institution{Outbrain}
  \country{Israel}
  \city{Netanya}
}

\blfootnote{Authors' emails: jhartman@outbrain.com; aklein@outbrain.com; dkopic@outbrain.com; nsilberstein@outbrain.com.}

\begin{abstract}
In this work, we introduce the notion of \textit{Context-Based Prediction Models}. A Context-Based Prediction Model determines the probability of a user's action (such as a click or a conversion) solely by relying on user and contextual features, without considering any specific features of the item itself. 
We have identified numerous valuable applications for this modeling approach, including training an auxiliary context-based model to estimate click probability and incorporating its prediction as a feature in CTR prediction models.
Our experiments indicate that this enhancement brings significant improvements in offline and online business metrics while having minimal impact on the cost of serving. 
Overall, our work offers a simple and scalable, yet powerful approach for enhancing the performance of large-scale commercial recommender systems, with broad implications for the field of personalized recommendations.
\end{abstract}



\maketitle

\section{Introduction}

Click-through rate (CTR) prediction models are a critical component of online recommender systems, as they help estimate the likelihood that a user will click on a recommended item. In online advertising systems, the accuracy of these models is crucial for the success of advertising campaigns and the revenue generated by publishers. Advertisers rely on CTR prediction models to target their ads to the right audience and optimize their advertising budget, while publishers use these models to maximize their revenue by displaying ads that are most likely to be clicked. 
CTR prediction techniques continue to be an active area of research in both industry and academia~\cite{mcmahan2013ad, he2014practical,anil2022factory}. In many commercial use cases, the CTR prediction model consists of billions of weights and must perform inference billions of times per second~\cite{hartman2021scaling}. Therefore, any improvements applied to the model must be carefully balanced with the cost of serving. 

In this paper, we introduce the notion of \textit{Context-Based Prediction Models} and demonstrate its effectiveness. A Context-Based Prediction Model determines the likelihood of an action (such as a click or a conversion) by solely considering user and contextual features, without taking into account any specific characteristics of the item itself. We have identified numerous valuable applications for this approach, including training a context-based model to estimate click probability and incorporating its predictions as features in our CTR prediction models. We demonstrate its value in two different use cases in the online advertising domain, specifically real-time bidding and ad recommendation.

With our implementation of contextual modeling, we effectively elevated both offline and online performances of our CTR models. The accompanying rise in the cost of serving was minimal in comparison to the substantial improvements witnessed in business metrics. Our models now utilize this improvement in production.

\section{Motivation}

In this paper, we aim to show the usefulness of context-based prediction models as observed at Outbrain -- a content recommendation and online advertising company, which operates several large-scale recommender systems to power the recommendations for the open web. 

The need for a context-based prediction model arose from the fact that our supply (webpages and widgets in which we show recommendations) has a high variance of attractiveness to advertisers. For instance, a placement at the top of the page is often more attractive than one at the bottom; or a user with a rich browsing history and interests profile, compared to a new user. We were interested in quantifying this variance and finding ways to adjust the pricing strategy for the different assets. Given our extensive predictive modelling capabilities, computing context-based predictions proved to be a simple but effective way to achieve our goals. 

A natural property of a useful feature is that it starts becoming widely used, and as such we are computing context-based predictions in two separate stacks, inside our core ad recommender systems and our real-time bidder~\cite{wang2017display}. We then use it to support several systems, including: as an auxiliary signal to the main click-through rate, conversion rate, and other prediction models; to direct more computational resources to recommendation requests which have a higher probability of being interacted with; to better segment our supply for interaction performance; to enable model-based exploration of our supply~\cite{hartman2022exploration}; and more. While context-based prediction has many uses, in this paper we focus on the benefits we observed by leveraging it as an auxiliary signal in our main CTR prediction models.

\section{Implementation}
\label{sec:Implementation}

Constructing a context-based prediction model could be tackled in a multitude of ways. We decided that one of the key requirements for it was reusability: the same context-based predictions should be usable in different use cases (described in the previous section). Since one of the primary goals was to improve existing CTR prediction models, we chose to construct the context-based prediction model as an auxiliary model. This model predicts context CTR and is entirely separate -- the main CTR model then uses contextual CTR predictions as a feature. 

Having a separate model for context CTR prediction has several additional benefits. Simplicity is essential as the contextual CTR model can use the same learning and prediction process as existing models, making it straightforward to construct and maintain in production. The fact that models are independent also makes it easier to test or swap each model separately. Lastly, in a large-scale recommender system, scaling separate models is also much less challenging.

Due to the sheer scale of billions of predictions per second, terabytes of data daily, and low latency requirements, any predictive models we use must scale well, i.e. be very compute-efficient. They must also be able to train online and incrementally. In order to tackle tasks like CTR prediction, Outbrain uses machine learning algorithms from the factorization machine~\cite{rendle2010factorization} (FM) family such as field-aware FM~\cite{juan2016field} and deep equivalents like DeepFM~\cite{guo2017deepfm} or Deep\&Cross V2~\cite{wang2021dcn}. These algorithms efficiently handle large and sparse datasets and excel at modeling feature interactions in such datasets.

Computing billions of predictions every second takes a significant amount of compute resources, thus we highly prefer improvements to models that do not increase their complexity. An interesting observation is that even though the main CTR prediction models have access to all data that context models have access to, we still observe significant prediction improvements when we use context CTR as an auxiliary signal. With neural network-based models, we can often improve the model simply by enlarging the network and thus increasing the model's capacity~\cite{hornik1989multilayer}. With enough capacity, the network should be able to model problems like context CTR with a subnetwork, thus rendering our approach of using a separate model moot. However, using such large models is infeasible under compute and latency constraints like ours.

Using an additional model in serving naturally incurs extra compute costs and an increase in latency. However, these downsides are not significant for a few reasons. During serving, the context CTR model computes far fewer predictions than the main CTR model. Because it relies only on the context, we only need to compute it once per request and can reuse it for all scored ads. Furthermore, this model can be simpler and thus more lightweight due to a smaller selection of features that are available for it to use. Finally, as we demonstrate in the next section, using the auxiliary model's predictions enables us to extract more information from the same data and thus considerably slim down the main model, meaning that we reduce the compute costs and latency overall.

\section{Results}
\subsection{Offline Evaluation}
\label{sec:offline}
As mentioned in Section~\ref{sec:Implementation}, our system employs a variety of algorithms for CTR prediction. In the following, we focus on two of them, namely, FFM~\cite{juan2016field} and Deep\&Cross V2~\cite{wang2021dcn}, utilized within our ad recommender systems and real-time bidder, respectively. Additionally, we explore two distinct approaches for incorporating context CTR predictions into these models.

The first approach is simply to replace a subset of features that are already included in the auxiliary model for the context CTR prediction, namely, introducing one new feature in our CTR prediction models and removing all the redundant ones. 
The second approach is to add the new context CTR prediction feature on top of the existing features, allowing some information redundancy in our models. 

For offline evaluation of the proposed approaches, we conduct experiments 
using hundreds of millions of impressions from logged Outbrain data. 
We compare the performance of the models with and without context CTR predictions by using Relative Information Gain (RIG)~\cite{metrics}, a linear transformation of log-loss given by \[RIG = 1-\frac{-c\cdot log(p)-(1-c)log(1-p)}{-\gamma\cdot log(\gamma)-(1-\gamma)log(1-\gamma)},\]
where $c$ and $p$ are an observed click and a click prediction, respectively, and $\gamma$ is the CTR of the evaluation data.
In addition, we consider the impact of the number of features in a model on the computation complexity 
of predicting, presented with an approximation of their floating-point operations (FLOPs).
The results are summarized in Table~\ref{tab:offline}.
\begin{table}[h!]
\begin{tabular}{c|cc|cc}
\hline
 \multirow{2}{*}{Usage of context CTR} &
 \multicolumn{2}{c}{Ad recommendation} &
 \multicolumn{2}{c}{Real-time bidder} \\
 & RIG lift  & FLOPs change & RIG lift  & FLOPs change\\
\hline
Replacing existing contextual features & +0.65\%   &  -38.48\% & +2.33\%   &  -24.4\% \\
Adding on top of existing features &+1.78\%   &  +3.85\% & +5.58\%   &  +1.5\% \\
\hline
\end{tabular}
\caption{Offline evaluation: comparison to baseline~models.}\label{tab:offline}
\vspace{-0.3cm}
\end{table}

The provided table demonstrates that integrating context CTR as a new feature to a CTR prediction model results in a significant improvement of the offline metric. Additionally, when employing the approach that eliminates redundant features from the CTR model, we can see that computational costs (in FLOPs)\footnote{We employ FLOPs as a proxy to estimate serving costs.} are reduced, showing a trade-off between improved model performance and computational efficiency. 
Therefore, we have the flexibility to select a suitable trade-off point that optimizes the desired metric, i.e. model quality or computation costs.

\subsection{Online Evaluation}

\begin{figure}[h]
  \centering  \includegraphics[width=0.5\columnwidth]{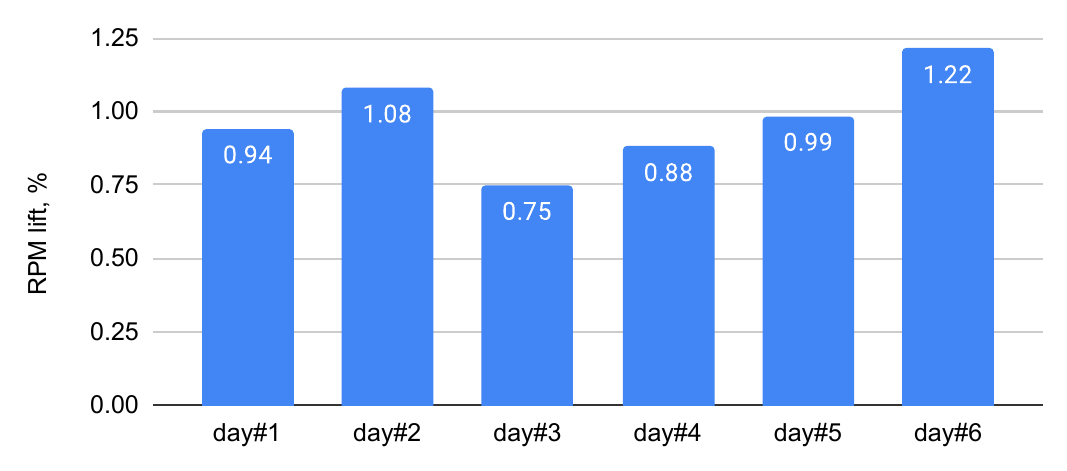}
  \caption{Daily RPM lifts of the variant including context CTR prediction as a new feature. }
  \label{fig:FFM rpm}
\end{figure}

In this section, we present the online evaluation of the CTR prediction model for our ad recommendations that incorporates the new context CTR feature. We choose to proceed with the model that demonstrates higher offline lifts, in other words, the model with some features already present in the auxiliary model (see Section~\ref{sec:offline}).
Consequently, a new model that is identical to the current production model with the addition of the new context CTR feature was trained using a few months' worth of historic logged data. Then it was tested in our online A/B testing system, serving a portion of Outbrain production traffic. 
The primary performance metric we utilize is revenue per thousand impressions (RPM). 
The daily RPM lifts over a few days are presented in Fig.~\ref{fig:FFM rpm}. The figure depicts consistent RPM lifts, averaging at +0.97\% over the six-day period showcased~\footnote{We also observed lifts in the real-time bidder use case.}.

\vspace{-0.5cm}

\section{Conclusion}

In this paper, we introduced the concept of Context-Based Prediction Models for enhancing large-scale recommender systems. Through two distinct use cases (real-time bidding and ad recommendations), we showcased the effectiveness of this approach, revealing substantial enhancements in both offline and online, business metrics. 
The context-based predictions are highly useful for several other downstream tasks in our system, e.g. throttling, resource allocation, and model-based exploration.
As a result, we believe that this concept possesses significant value and can be successfully applied in other domains.

\section*{Speaker Bio}

\textbf{Jan Hartman} is a data scientist/machine learning engineer at Outbrain, where he works with high-throughput, low-latency ML pipelines at a large scale. He tackles implementing and applying state-of-the-art models for click prediction. Before joining Outbrain, he worked on research projects in the fields of distributed computing, neural network optimization, and cryptography. 
Honors MSc degree in Computer \& Data Science from the University of~Ljubljana. Open-source contributor. Research interests include deep learning, neural network embeddings, and model compression. 

\textbf{Dr. Natalia Silberstein} is a senior data scientist and team leader in the Recommendations group at Outbrain. Her responsibilities involve enhancing and developing algorithms for personalized ad selection. Prior to joining Outbrain, she worked as a research scientist at Yahoo Research Haifa in the Native Ad Science group. Before that, Natalia was part of the Mail Mining group, where she primarily worked on analyzing and modeling mail data to devise novel mail features. 
She holds a PhD from the Computer Science Department at the Technion - Israel Institute of Technology. After completing her doctoral studies, she conducted postdoctoral research at the Department of Electrical \& Computer Engineering, University of Texas, Austin, focusing on coding for distributed storage systems.


\begin{acks}
We would like to thank 
Yulia Stolin, Andra\v{z} Tori, Jure Ferle\v{z}, Robert Dov\v{z}an, Chen Weiss, Moran Haham, Or Shoham, Gal Giladi Levi, Danny Kidron, and Yonatan Zusman
for their valuable contributions to this project.
\end{acks}

\bibliographystyle{ACM-Reference-Format}
\bibliography{bibliography}


\end{document}